\documentclass{svproc}
\usepackage{url}
\usepackage{makeidx}
\usepackage{graphicx} 
\usepackage{newtxmath}
\usepackage[bottom]{footmisc}
\usepackage{newtxtext} 

\usepackage{color}

\begin{document}
	\mainmatter              
	\title{Combination Therapy for Chronic Hepatitis B Using Capsid Recycling Inhibitor }
	\titlerunning{Combination Therapy for Chronic Hepatitis B Using Capsid Recycling Inhibitor }  
	%
	\author{\underline{Rupchand Sutradhar}\inst{1}[0000-0002-8444-2150]  \and D C Dalal\inst{1} [0000-0001-6537-5282]}
	\authorrunning{Sutradhar and Dalal} 
	%
	\tocauthor{Rupchand Sutradhar, D C Dalal}
	\institute{Indian Institute of Technology Guwahati, Guwahati, Assam, 781039, India\\
		\email{rsutradhar@iitg.ac.in} (\underline{Rupchand Sutradhar}),\\ WWW home page:
		\texttt{https://sites.google.com/view/rupchand/home}\\
		\email{durga@iitg.ac.in} (D C Dalal)\\
		\email{rsutradhar@iitg.ac.in} ($^*$corresponding author)
	}
	
	\maketitle              
	
	\begin{abstract}
		 In this paper, we investigate the dynamics of hepatitis B virus infection taking into account the implementation of combination therapy through mathematical modeling. 
		This model is established  considering the interplay between uninfected cells, infected cells, capsids, and viruses.   Three drugs are considered for specific roles: (i) pegylated interferon (PEG-IFN) for immune modulation, (ii) lamivudine (LMV) as a reverse-transcriptase inhibitor, and (iii) entecavir (ETV) to block capsid recycling.
		Using these drugs, three combination therapies are introduced, specifically CT: PEG-IFN+LMV, CT: PEG-IFN+ETV, and CT: PEG-IFN+LMV+ETV. 
		As a result, when LMV is used in combination therapy with PEG-IFN and ETV, the impacts of ETV become insignificant. In conclusion, if the appropriate drug effectively inhibits reverse-transcription, there's no need for an additional inhibitor to block capsid recycling.
	\end{abstract}
	Abbreviations: HBV: Hepatitis B virus, CT: Combination therapy, MT: Monotherapy, PEG-IFN: Pegylated interferon, LMV: Lamivudine, ETV: Entecavir, CT: PEG-IFN+LMV: Combination therapy with PEG-IFN and LMV, CT: PEG-IFN+ETV: Combination therapy with PEG-IFN and ETV, CT: PEG-IFN+LMV+ETV: Combination therapy with PEG-IFN, LMV and ETV. 
	\section{Introduction}
	Hepatitis B virus (HBV) acts as an etiological agent of the  life-threatening viral infection, hepatitis B. HBV infection is a continuing major public health problem with global implications. Indeed, despite the presence of a preventive vaccine for this virus, approximately 250 million people worldwide are  suffering from chronic HBV infection, which stands as a primary contributor to cause advanced liver diseases, such as cirrhosis, hepatocellular carcinoma. It is also estimated that nearly 1.5 million new cases were found each year \cite{WHO_2023}. This grievous viral  infection is generally transmitted vertically as well as horizontally. There are mainly two routes the viral infection can spread: virus-to-cell and cell-to-cell transmission.
 HBV infection can be categorized into (a) acute: which is a short-term infection last less than six months, and (b) chronic: which persists over the long term. 
	
	Currently, two therapeutic approaches are available for treating HBV-infected patients: (i) nucleotide analogues, and (ii) immune modulator drugs (standard or pegylated interferon alpha-2a, alpha-2b) \cite{2021_philips_critical,2009_loomba}. Among the oral antiviral drugs, tenofovir disoproxil, tenofovir alafenamide, and entecavir are considered as first-line treatments, while telbivudine, adefovir dipivoxil, and lamivudine are categorized as second-line options.
	Antiviral medications are designed to retard or block the virus reproduction  whereas the  immune modulator drugs enhance the immune system's capabilities to aid in elimination of viruses. The choice between monotherapy  and combination therapy (CT)  in the treatment of HBV infection depends on several factors, including the patient's disease stage, viral load, liver function, etc. One of the major limitations of  monotherapy is its vulnerability to the development of drug-resistant strains. Moreover, monotherapy is not enough to suppress the virus completely, and there is a possibility of risk of relapsing the viral infection after discontinuation of therapy. On the other hand, CT can reduce  the risk of  resistance, enhance the  efficacy, lead to  more comprehensive and effective viral suppression in compared with monotherapy because in  CT, different drugs  target different stages of the HBV life cycle. Despite the benefits of CT, it's important to acknowledge the significant challenges it presents, including drug-drug interactions, a higher pill burden, increased risk of toxicity and increased treatment costs, etc. For the patients with high viral load, significant liver disease, or  with a history of resistance to a particular monotherapy, CT is often recommended. Paul and Han \cite{2011_Paul} analyzed the current roles of CT in chronic hepatitis B infection and mentioned several CTs in various settings. Hui et al. \cite{2008_Hui_96 week} conducted a study comparing the combination of adefovir and emtricitabine to adefovir alone in a cohort of 30 patients over 96 weeks. In that study, Hui et al. \cite{2008_Hui_96 week}  observed that while the combination group exhibited significantly higher levels of HBV DNA suppression and ALT normalization, the rate of HBeAg seroconversion remained similar in both groups. The utilization of multi-target therapies for treating complex diseases can increasingly benefit patients \cite{2022_mello}. It is noticed that patients with chronic hepatitis B who were treated with a combination of telbivudine and tenofovir showed significant improvements in both the virus levels and biochemical markers over a one-year period of time \cite{2013_Panda}. Recently, Ntaganda \cite{2021_Ntaganda} investigated the impacts of combination therapies on chronic hepatitis B virus infection using Pontryagin's maximum principle and concluded that employing a combination of two drugs can effectively control the HBV infection, promoting a healthy life.

In the last couple of decades, numerous mathematical models have been proposed  to investigate the dynamics of various types of viral infection, such as HBV, Covid-19, HIV \cite{2007_Dahari,2018_danane,2018_chenar,2022_ICNDA_Banerjee_Saha}. In case of HBV infection, Nowak et al. \cite{1996_Nowak} first proposed a HBV infection dynamics model in the year 1996. The model was proposed by Nowak et al. is commonly known as basic model. This basic model comprises three compartments, including susceptible host cells, infected cells, and free virus particles. Following the pioneering work of Nowak et al., extensive research has been carried out modifying the basic model or formulating  new dynamics model based on the biological findings available in the literature \cite{2018_chenar,2006_Murray,2007_Ciupe,2008_Min}.   These models are useful in understanding the pathogenesis of infection as well as in  advancing the knowledge about the virus and disease.
		 Fatehi et al. \cite{2018_chenar} determined that in HBV infection, NK cells take a significant role in apoptosis as it can kill the infected cells by producing the proteins perforin and granzymes. Murray et al. \cite{2006_Murray} proposed another HBV infection dynamics model with three compartments (infected cells, capsids and viruses) and measured that  the half-life of HBV is approximately four hours.
Besides, we have recently proposed an advanced HBV infection dynamics model that underscores the significance of  recycling of capsids \cite{2023_Sutradhar}.
	In the present study, we have extended our previously proposed model mentioned in \cite{2023_Sutradhar} with some symbolic changes by including the effects of antiviral combination therapies.
	This new model considers three different drugs (PEG-IFN, LMV, and ETV) in its development.

	The  details of the drugs used in the combination therapy  are given below.
	\begin{enumerate}
		\item\textbf{Immune modulator drug (PEG-IFN): }
		IFNs function by interacting with IFN receptors on the cell membrane, setting off a chain reaction of secondary messenger activation. Through multiple complex mechanisms, IFNs prevent the infected hepatocytes, active the natural killer cells and other immune cells, increase the production of cytokines, suppress the viral protein synthesis, degrade the viral mRNAs and enhance the antigen presentation by human leukocyte antigen (HLA) I and II to the immune system \cite{2008_Dusheiko}. IFNs offer a significant advantage over nucleoside analogues in terms of the potential for immune-mediated viral clearance and in preventing the development of viral resistance.
		
		\item  \textbf{Reverse-transcriptase inhibitor (LMV):} 
		LMV was initially developed as an inhibitor for HIV reverse-transcriptase for the treatment of AIDS patients. It was subsequently approved for HBV by the United States Food and Drug Administration in the year 1998  as it also effectively inhibits  HBV reverse-transcriptase  \cite{2018_Quercia}. LMV is a dideoxynucleoside cytosine analogue that can exert its antiviral effects by inhibiting viral DNA synthesis through reverse-transcriptase DNA chain termination, a process that occurs following phosphorylation. 
		\item \textbf{Capsid recycling inhibitor (ETV): } ETV is a cyclopentyl guanosine nucleoside analogue. It was granted approval for use in the United States in the year 2005.  By preventing the rcDNA formation, ETV  can inhibit the  intracellular recycling of this viral  genome in the cytoplasm which results in potent inhibition of HBV  DNA polymerase \cite{2021_Ohsaki}.
	\end{enumerate}
	Using these three above-mentioned drugs, we set three different CTs: (i) CT: PEG-IFN+LMV, CT: PEG-IFN+ETV and CT: PEG-IFN+LMV+ETV. 
	In order to manage the HBV infection, Paul and Han \cite{2011_Paul} discussed various combination therapies with currently available drugs including their efficacy and safety profiles. Before implementing the aforementioned combination therapies, the compatibility of the patients with these CTs is ensured by the study conducted by Paul and Han \cite{2011_Paul}. Three in silico experiments have been performed subsequently. The experiment results of this study highlight the potential benefits of antiviral therapies in the case of HBV infection.
	
	\section{Model formulation with combination therapy}
	Neglecting the effects of  drug-drug interaction and drug resistance strain,  the temporal change  of each component of the proposed model  are derived as follows:
	\begin{equation}
		\left.
		\begin{split} \label{main model}
			&\dfrac{dT}{dt}=\lambda-\delta_T T-(1-\epsilon_1)kVT=g_1(T,I,C,V),\\
			&\dfrac{dI}{dt}=(1-\epsilon_1)kVT-\delta_I I=g_2(T,I,C,V),\\
			&\dfrac{dC}{dt}=a(1-\epsilon_2)I+\gamma(1-\epsilon_3)(1-\alpha)C-\alpha\beta C-\delta_C C=g_3(T,I,C,V),\\
			&\dfrac{dV}{dt}=\alpha \beta C-\delta_V V=g_4(T,I,C,V). 
		\end{split} 
		\right\}
	\end{equation}
	 Here, $T,~I,~C$ and $V$ are functions of time only  and represent uninfected hepatocytes, infected hepatocytes, HBV capsids and viruses, respectively. The corresponding  natural death rates of these four compartments ($T,~I,~C,~V$) are denoted by $\delta_T,~\delta_I,~\delta_C,$ and $\delta_V$. The uninfected cells ($T$) are produced with an average growth rate $\lambda$ and become infected with a rate $k$
	by the free virus. In this study, only virus-mediated infection are taken into account. It is considered that HBV DNA-containing capsids ($C$) are produced in two ways: $(i)$ from the infected hepatocytes with a production rate $a$, and $(ii)$  through  recycling of newly produced capsids  with recycling rate $\gamma$. It is thought that $\alpha$  volume fraction of newly produced capsids is transformed into infectious virus particles with a rate $\beta$, while the remaining part of the capsids is returned into the nucleus to replenish the pool of covalently closed circular DNA.  In this model, the impacts of three drugs are integrated. The first one is PEG-IFN which inhibits the new infections of healthy hepatocytes in the liver. The primary function of the second drug, LMV,  is to inhibit or stop the production of the virus by inhibiting reverse-transcription. As the third therapeutic agent, we incorporate ETV which inhibits  the  intracellular recycling of HBV DNA-containing capsids in the cytoplasm \cite{2021_Ohsaki}. The parameters $\epsilon_1,~\epsilon_2,$ and $\epsilon_3$ quantify the efficiency of PEG IFN, LMV and ETV, respectively.
	
	The right-hand side of each equation of the system \eqref{main model},  \textit{i.e.}, the functions $g_1,~g_2,~g_3$ and $g_4$  are polynomial functions of $T,~I,~C$, and $V$. Hence, each function is continuous and meets Lipschitz's condition on any closed interval $[0, \tau]$, where $\tau$ belongs to the set of  positive real numbers, denoted by $\mathbb{R^+}$. Consequently, the solution of the system \eqref{main model} exists and is unique for all initial conditions. The non-negativity and boundedness of the solution follow from the recent work conducted by Sutradhar and Dalal \cite{2023_Sutradhar} when the threshold $R_s:=\alpha\beta-(1-\alpha)(1-\epsilon_3)\gamma+\delta_C>0$. The biological interpretation of this threshold ($R_s$) is well-explained in the article of Sutradhar and Dalal  \cite{2023_Sutradhar_fractional}. The condition $R_s > 0$ will remain in effect for the rest of this study.   The basic reproduction number of this model \eqref{main model} is calculated using the well-known next-generation approach, and it is given by 
	$$R_0=\dfrac{k \lambda   a \alpha  \beta \left(1-\epsilon _1\right)(1-\epsilon_2)}{\mu  \left(\alpha  \beta   +\alpha   \gamma   (1-\epsilon _3)- \gamma  (1-\epsilon _3) + \delta_C \right)\delta_I\delta_V}.$$
	It's important to note that we do not present the experimental validation as a separate component of this study as Sutradhar and Dalal \cite{2023_Sutradhar}  have already validated this model \eqref{main model} without therapy  with the experimental data of chimpanzees. 
	
	\section{Steady-states and their stabilities}
	In order to determine the steady-states of  the system \eqref{main model}, we solve the system of equations $g_1=0,~g_2=0,~g_3=0$ and $g_4=0$. As a result, it is seen that the  system \eqref{main model} has the following infection-free steady-state $E_0=\left(\dfrac{\lambda }{\delta_T }, 0, 0, 0\right)$ and a unique endemic steady-state $E_i=(T^*,~I^*,~C^*,~V^*)$, where 
	
	\begin{align*}
		T^*&=\dfrac{\delta _I \delta _V \left(\alpha  \beta +\epsilon _3 (\gamma -\alpha  \gamma )+(\alpha -1) \gamma +\delta _C\right)}{a \alpha  \beta  k \left(\epsilon _1-1\right) \left(\epsilon _2-1\right)},\\
		I^*&=\dfrac{a \alpha  \beta  k \lambda  \left(\epsilon _1-1\right) \left(\epsilon _2-1\right)-\delta _I \delta _T \delta _V \left(\alpha  \beta -(\alpha -1) \gamma  \epsilon _3+(\alpha -1) \gamma +\delta _C\right)}{a \alpha  \beta  \delta _I k \left(\epsilon _1-1\right) \left(\epsilon _2-1\right)},\\
		C^*&=\frac{a \lambda  \left(1-\epsilon _2\right)}{\delta _1 \left(\alpha  \beta -(\alpha -1) \gamma  \epsilon _3+(\alpha -1) \gamma +\delta _C\right)}+\frac{\delta _T \delta _V}{\alpha  \beta  k \epsilon _1-\alpha  \beta  k},\\
		V^*&= \frac{a \alpha  \beta  \lambda  \left(1-\epsilon _2\right)}{\delta _1 \delta _V \left(\alpha  \beta -(\alpha -1) \gamma  \epsilon _3+(\alpha -1) \gamma +\delta _C\right)}+\frac{\delta _T}{k \left(\epsilon _1-1\right)}.
	\end{align*}
	In the context of biology, the endemic steady-state exists  whenever $R_s>0$ and $R_0>1$. Considering two suitable Lyapunov functionals $\mathbb{L}_{1}(t)$ and $\mathbb{L}_{2}(t)$ given below, the global asymptotic stability  of both  steady-states are established applying Lyapunov-LaSalle invariance principle.
	\begin{align*}
		\mathbb{L}_{1}(t)&=\dfrac{\lambda}{\delta_T}\left\{\frac{\delta_T T(t)}{\lambda}-1-\ln\left (\frac{\delta_T T(t)}{\lambda}\right )\right\}+I(t)+\frac{\delta_I}{a(1-\epsilon_2)}C(t)\\
		&+\frac{\delta_I(\alpha\beta-\gamma(1-\epsilon_3)(1-\alpha)+\delta_C)}{a(1-\epsilon_2)\alpha\beta}V(t).\\	
		\mathbb{L}_{2}(t)&=T^*\left\{\frac{T(t)}{T^*}-1-\ln\left (\frac{T(t)}{T^*}\right )\right\}+I^*\left\{\frac{I(t)}{I^*}-1-\ln\left (\frac{I(t)}{I^*}\right )\right\}\\
		&+\frac{\delta_I C^*}{a(1-\epsilon_2)}\left\{\frac{C(t)}{C^*}-1-\ln\left (\frac{C(t)}{C^*}\right )\right\}\\
		&+\frac{\delta_I (\alpha\beta-(1-\alpha)(1-\epsilon_3)\gamma+\delta_c) V^*}{a(1-\epsilon_2)\alpha\beta}\left\{\frac{V(t)}{V^*}-1-\ln\left (\frac{V(t)}{V^*}\right )\right\}. 
	\end{align*}
	In terms of stability, it is proved that
	\begin{enumerate}
		\item When $R_0<1$, the disease-free steady-state becomes globally asymptotically stable.
		\item In the case of $R_0>1$, the endemic steady-state achieves global asymptotic stability.	
	\end{enumerate}	
	\section{Numerical simulation } As  the proposed model is highly non-linear in nature, the  system \eqref{main model} is solved numerically. The values of model parameters with slight symbolic modifications (except $\epsilon_1, \epsilon_2, \epsilon_3$)  are predominantly adopted from the most recent study conducted by Sutradhar and Dalal \cite{2023_Sutradhar}. The mean antiviral
	effectiveness of PEG-IFN 100/200g monotherapy is 82.6\%, \textit{i.e.}, the value of $\epsilon_1=0.826$ \cite{2005_Sypsa}.  According to the work of Dahari et al. \cite{2007_Dahari},
	the treatment in blocking the virion production by reverse-transcriptase inhibitors is nearly 95\% effective. Therefore, the value of $\epsilon_2$  is considered  0.95. The inhibition rate of ETV, $\epsilon_3=0.89$,  is taken from the article of Kitagawa et al. \cite{2023_Kitagawa}. 
	
	In this study, three experiments are conducted using the above-mentioned combination therapies. In all three experiments, the drugs are administered during chronic infection, and the simulation results are presented below.  The initial conditions in each of the three cases, are characterized as the chronic stages of the patients.  
	\subsection{Experiment-1: The effects of CT: PEG-IFN+LMV ($\epsilon_1\neq0,~\epsilon_2\neq 0,~\epsilon_3=0$)}
\begin{figure}[h]
		\centering
		\includegraphics[width=12cm,height=9cm]{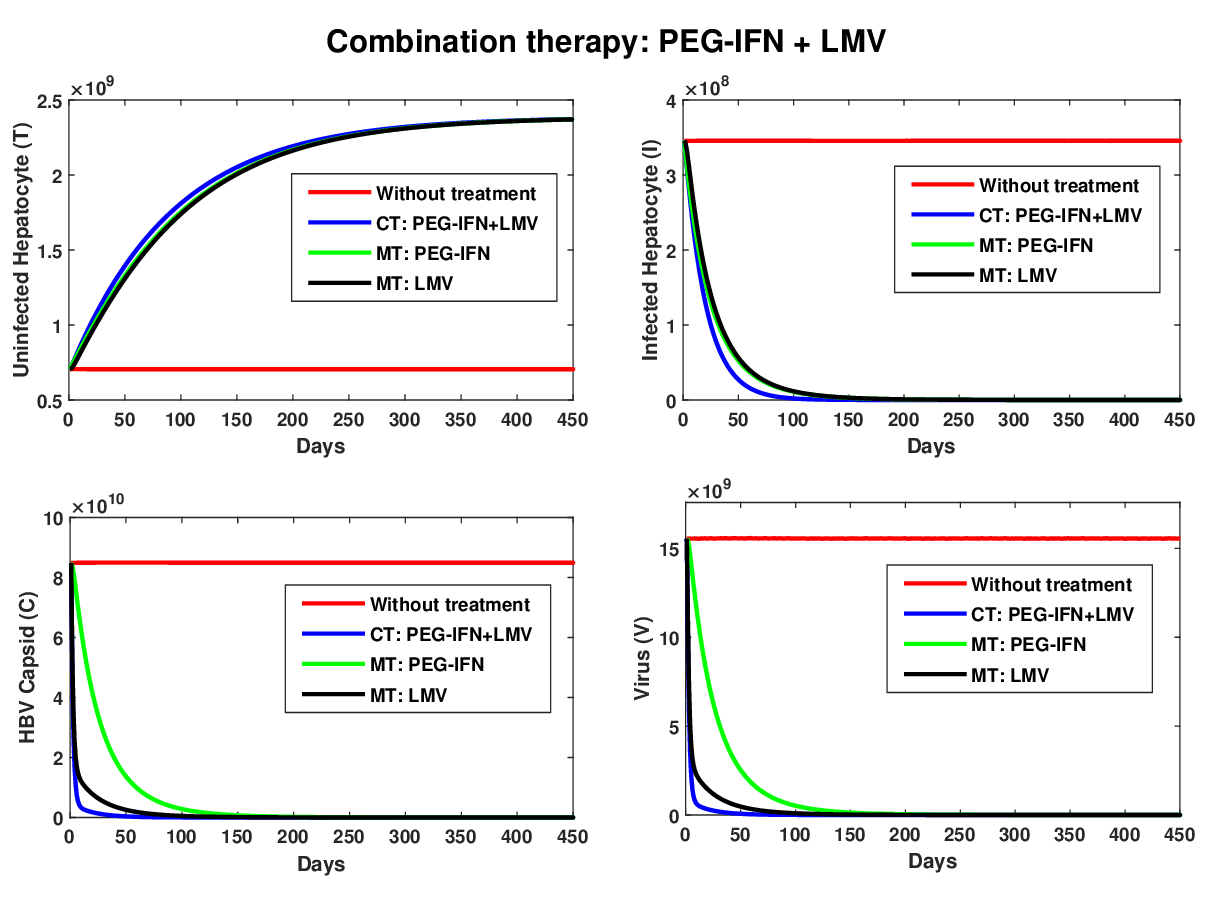};
		\caption{The effects of therapy are shown here. The solutions of the system  \eqref{main model} are described by (i) Red line: solution without treatment, (ii)  Blue line: solution of the system under the CT with PEG-IFN and LMV, (iii) Green line: solution of the system under MT with PEG-IFN, and (iv)  Black line: solution of the system under MT with LMV.}
		\label{Treatment: PEG IFN+LMV}
	\end{figure}

	In this experiment, CT involving two drugs, PEG-IFN and LMV, is applied to the specific location within the dynamical system \eqref{main model}. The value of $\epsilon_3$ (efficiency of ETV) is considered zero in this case. 
	Fig.\ref{Treatment: PEG IFN+LMV} shows the evolution of uninfected hepatocytes, infected hepatocytes, capsids and viruses against time (days) presenting the following facts:
	\begin{itemize}
		\item Combination therapy with PEG-IFN and LMV.
		\item Monotherapy with PEG-IFN.
		\item Monotherapy with LMV.
	\end{itemize} 
\noindent
 Combination therapy of  PEG-IFN with the nucleos(t)ide analog LMV leads to a significantly higher rate of HBeAg seroconversion compared to using either LMV or PEG-IFN monotherapy. Although the population of uninfected cells with therapy is higher than that observed in the case without any therapies (CT and mono).  The inclusion of PEG-IFN to the combination therapy regimen results in enhancement of effectiveness  due to the additive effects of CT. It is also observed that  the results of this CT show a rapid recovery from the infection compared to the monotherapies. Recently, Su and Liu \cite{2017_Su} arrived at similar conclusions (as seen in 	Fig.\ref{Treatment: PEG IFN+LMV}) concerning the effectiveness of this combination therapy. So, in conclusion, the CT with PEG-IFN and LMV is better than the mono therapy with PEG-IFN or LMV.
	\subsection{Experiment-2: The effects of CT: PEG-IFN+ETV ($\epsilon_1\neq0,~\epsilon_2=0,~\epsilon_3\neq0$)}
\begin{figure}[h]
		\centering
		\includegraphics[width=12cm,height=9cm]{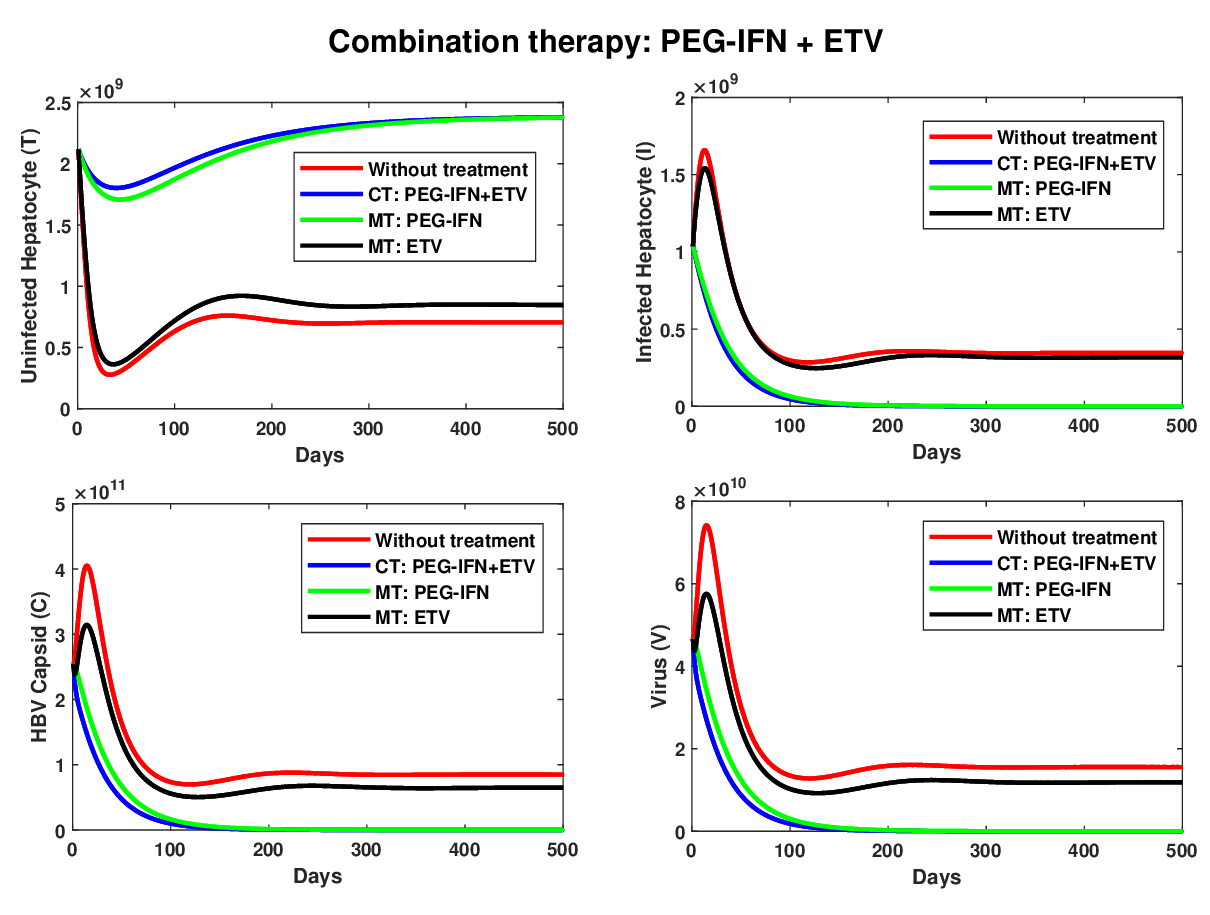};
		\caption{The effects of  therapy are demonstrated here. The solutions of the system  \eqref{main model} are described by (i) Red line: solution without treatment, (ii) Blue line: solution of the system under the CT with PEG-IFN and ETV, (iii) Green line: solution of the system under MT with PEG-IFN, and (iv) Black line: solution of the system under MT with ETV.}
		\label{Treatment: PEG IFN+ETV}
	\end{figure}

	Fig. \ref{Treatment: PEG IFN+ETV} displays the dynamical variations in all four compartments ($T,I,C,V$) over time, showcasing  the following scenarios:
		\begin{itemize}
		\item The effects of combination therapy with PEG-IFN and ETV.
		\item The effects of monotherapy with PEG-IFN.
		\item The effects of monotherapy with ETV.
\end{itemize} 
	  In this experiment, the  nucleos(t)ide ETV is applied as  capsids recycling inhibitor. Although, its primary mode of action involves the inhibition of reverse-transcription which is an essential step in viral DNA replication.  The solutions of the system \eqref{main model} in the absence of such treatment are compared with the corresponding solution of the system without control. The administration of these medications has led to the following documented outcomes:
	\begin{enumerate}
		\item  Monotherapy with ETV does not appear to be highly effective in eradicating the infection completely but could reduce the concentration of capsids as well as viruses \cite{2022_Leowattana}.  Therefore, relying solely on the inhibiting capsid recycling seems not to be an effective antiviral strategy. It is also observed that the efficacy of treatment of PEG-IFN monotherapy is more pronounced  compared to ETV monotherapy.
	   \item The use of a CT involving PEG-IFN and ETV  has demonstrated success in achieving a complete sterilizing cure of the viral infection and has also shown a continued clinical benefit.
	\end{enumerate}
	\subsection{Experiment-3: The effects of CT: PEG-IFN+LMV+ETV ($\epsilon_1\neq0,~\epsilon_2\neq0,~\epsilon_3\neq0$)}
\begin{figure}[h!]
		\centering
		\includegraphics[width=12cm,height=9cm]{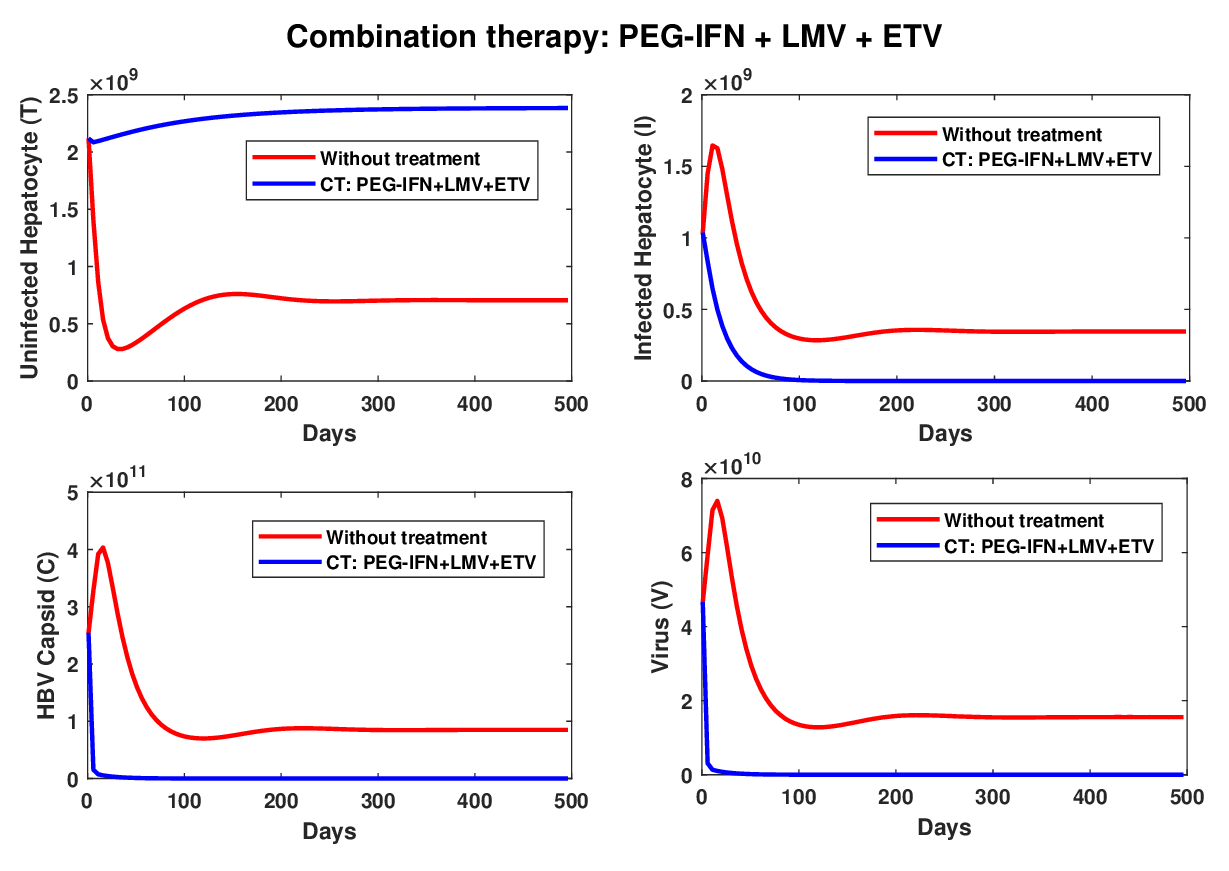};
		\caption{The effects of combination therapy with three drugs (PEG-IFN+LMV+ETV) are shown here. The solution of the system \eqref{main model} without treatment is illustrated by the red curves, while the blue curves depict the solution of the system \eqref{main model} under treatment.}
		\label{Treatment: PEG IFN+LMV+ETV}
	\end{figure}

	The simultaneous application of more than two drugs to treat a particular disease is rarely documented in the literature. To the based on our knowledge,  the CT: PEG-IFN+LMV+ETV is likely studied for the first time. In this case, out of the three drugs, two of them (LMV and ETV) are employed to inhibit the viral replication step, while the third one (PEG-IFN) is applied to boost the host immune system. In Fig. \ref{Treatment: PEG IFN+LMV+ETV}, the outcomes of this treatment  are visualized.  The reduction in viral load begins immediately after the administration of these drugs, and finally leads to the patient of becoming infection-free. Due to the implementation of this CT, a substantial increase is observed in the profile of uninfected hepatocytes. On the other hand, significant reduction is noticed on  the number of infected hepatocytes, capsids and viruses. Upon the administration of this CT, the concentration of infected hepatocytes, capsids and viruses  disappear completely.  In the context of HBV infection, this CT has also shown to be a highly effective antiviral treatment but before implementing this therapy in practical use, it is essential to obtain clinical approval to assess and mitigate potential side effects and other complications.
	\subsection{A comparative examination of the treatments employed in Experiment-1, Experiment-2 and Experiment-3}
\begin{figure}[h!]
		\centering
		\includegraphics[width=12cm,height=9cm]{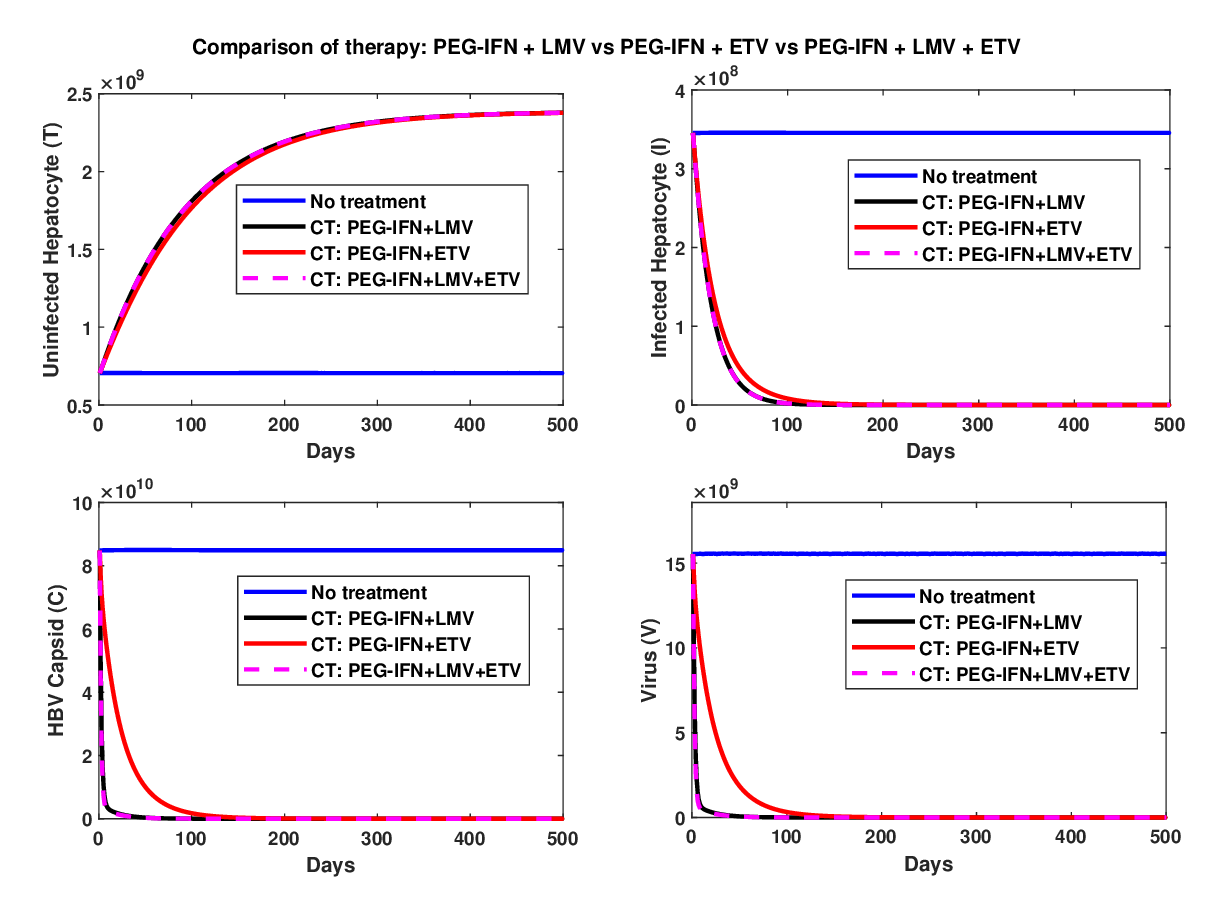}
		\caption{The comparison of different therapies. Solid blue line:  no treatment, Dash black line: CT: PEG-IFN+LMV, Solid red  line:  CT: PEG-IFN+ETV, Dash magenta  line: CT: PEG-IFN+LMV+ETV.}
		\label{Treatment: comparison}
	\end{figure}

	The main aim of this study is to investigate the effectiveness of the combination therapy involving capsid recycling inhibitors. This section entails a comparative analysis of the efficacy of different  CTs as mentioned above. Under the  same initial conditions and parameter values, we solve the system \eqref{main model} considering different CTs and graphically present the corresponding  solutions in Fig. \ref{Treatment: comparison}. It is seen that all three of these combination therapies successfully eliminate the infection, but there are some notable differences observed in their dynamics.  The effectiveness  of CT: PEG-IFN+LMV (black curves in Fig. \ref{Treatment: comparison}) is found to be almost same with that of CT: PEG-IFN+LMV+ETV (dash magenta curves in Fig.  \ref{Treatment: comparison}).  It means that in the presence of LMV (reverse-transcriptase inhibitor) in CT, the effects of ETV (capsid recycling inhibitor) are not significant. In other words, this study suggests that there is no need to apply an additional inhibitor for capsid recycling if the reverse-transcription is  effectively blocked by the corresponding inhibitors.
	
	In the context when ETV is used as a capsid recycling inhibitor, the simulation results for capsid and virus compartments suggest that the efficacy of CT: PEG-IFN+ETV is lower than that of CT: PEG-IFN+LMV. 
	This might be due to the coordinated action of various pharmaceutical agents of the drugs. This is evident from the fact that CT: PEG-IFN+LMV leads to a more rapid decrease in the concentration of infected hepatocytes,  HBV capsids and viral load compared to those of  CT: PEG-IFN+ETV. Therefore, it will be more beneficial to a  patient to use  reverse-transcription inhibitor with PEG-IFN rather than impede capsid recycling.

	\newpage
	\section{Conclusions}
	In this paper, we have examined an HBV infection model that incorporates intracellular recycling of HBV DNA-containing capsids and evaluated the impacts of different antiviral therapies. This model is defined by a set of four ordinary differential equations that delineate the associations between uninfected cells, infected cells, capsids and viruses. Three drugs are considered to perform specific roles: an immune system modulator (pegylated interferon), a reverse-transcriptase inhibitor (lamivudine), and an agent for blocking the recycling of HBV capsids (entecavir).  The effects of the combination therapy involving these three approved well-defined drugs are also discussed in this study. We have established the existence and uniqueness of the solution. By constructing two suitable Lyapunav functionals, the global stability of each steady-state is established.  Based on the results obtained from both analyses and numerical experiments, the following conclusions are drawn:
	\begin{enumerate}
		\item  In the presence of LMV in the combination therapy, the impacts of ETV are not noticeable but when ETV is administered  with PEG-IFN, this combination is capable of clearing the infection.
		\item   The effectiveness  of CT: PEG-IFN+LMV is found to be almost same with that of CT: PEG-IFN+LMV+ETV. 
		\item  CT: PEG-IFN+LMV  results in a more rapid reduction in the concentration of HBV capsids and viral load when compared with the outcomes of CT-PEG-IFN+ETV.
		\item The control strategies that target both reverse-transcription and capsid recycling simultaneously does not seem to be a favorable option. The inhibition of either reverse-transcription or capsid recycling seems to be sufficient for eliminating the infection. However, it is observed that restraining the  reverse-transcriptase is the preferred option over suppressing the  capsids recycling.
	\end{enumerate}
	Utilizing the insights derived from these study, the proposed model holds significant potential for its application in a diverse array of clinical scenarios.  Furthermore, it can be used as a valuable tool for  the development of novel pharmaceutical agents and treatment strategies. However, prior to the implementation of the obtained results practically, experimental confirmation is a prerequisite. It is also important to consider the cost of combination therapy.  
	%
	%
	\subsection*{Competing interests}
	\noindent
	The authors declare that they have no competing interests.
	\subsection*{Authors' contributions}
	\noindent 
	Both authors contributed equally.
	
\end{document}